\documentclass[twocolumn,showpacs,preprintnumbers,amsmath,amssymb,pra]{revtex4-1}

\usepackage{graphicx}
\usepackage{dcolumn}
\usepackage{bm}
\usepackage{color}

\begin{document}

\title{Dissipative quantum bifurcation machine:
\\
Quantum heating of coupled nonlinear oscillators}

\author{Hayato Goto,$^{1}$ Zhirong Lin,$^{2}$ and Yasunobu Nakamura$^{2,3}$}

\affiliation{$^{1}$Frontier Research Laboratory, 
Corporate Research \& Development Center, 
Toshiba Corporation, 1, Komukai-Toshiba-cho, 
Saiwai-ku, Kawasaki 212-8582, Japan \\
$^{2}$RIKEN Center for Emergent Matter Science (CEMS), Wako, Saitama 351-0198, Japan \\
$^{3}$Research Center for Advanced Science and Technology (RCAST), 
The University of Tokyo, Meguro-ku, Tokyo 153-8904, Japan}

\begin{abstract}
A network of driven nonlinear oscillators without dissipation 
has recently been proposed for solving combinatorial optimization problems 
via quantum adiabatic evolution through its bifurcation point. 
Here we investigate the behavior of the quantum bifurcation machine 
in the presence of dissipation. 
Our numerical study suggests that 
the output probability distribution of the dissipative quantum bifurcation machine is 
Boltzmann-like, where the energy in the Boltzmann distribution 
corresponds to the cost function of the optimization problem. 
We explain the Boltzmann distribution 
by generalizing the concept of \textit{quantum heating} in a single oscillator 
to the case of multiple coupled oscillators. 
The present result also suggests that such driven dissipative nonlinear oscillator networks 
can be applied to \textit{Boltzmann sampling}, 
which is used, e.g., for Boltzmann machine learning in the field of artificial intelligence.
\end{abstract}


\maketitle

\section{Introduction}

Recently, hardware devices designed for combinatorial optimization have attracted much attention. 
The most well-known example is the quantum annealer developed by D-Wave Systems \cite{Johnson2011a}.
The machines are based on quantum annealing or adiabatic quantum computation 
\cite{Kadowaki1998a,Farhi2001a,Santoro2002a,Das2008a} 
and are physically implemented with superconducting quantum bits (qubits). 
Classical annealers with semiconductor classical bits in CMOS devices have also been studied \cite{Yamaoka2016a}. 
Both are designed to find the ground state of the Ising model. 
Such Ising machines are useful in the sense that many combinatorial optimization problems 
can be transformed into the Ising problem \cite{Lucas2014a}. 

Another approach to the physical implementation of Ising machines is based on parametric oscillations, 
where two stable oscillating states of each parametric oscillator correspond to up and down spins
\cite{Wang2013a,Marandi2014a,Haribara2015a,McMahon2016a,Inagaki2016a,Mahboob2016a,Goto2016a,Nigg2017a,Puri2017a}. 
There are two major types of such Ising machines. 
The first type originally proposed in \cite{Wang2013a} uses 
a network of optical parametric oscillators (OPOs). 
The threshold of an OPO is determined by one-photon loss and 
its oscillating states are stabilized by two-photon loss. 
The coupling between two OPOs is implemented by mutual injection \cite{Wang2013a,Marandi2014a}
or measurement feedback
\cite{Haribara2015a,McMahon2016a,Inagaki2016a}. 
The coupling does not conserve the energy of the network and 
consequently is accompanied by dissipation. 
The second type originally proposed in \cite{Goto2016a} uses 
a network of nondissipative parametric oscillators with Kerr nonlinearity (KPOs). 
The threshold of a KPO is determined by one-photon detuning and 
its oscillating states are stabilized by the Kerr effect (nonlinear energy shift). 
The coupling between two KPOs is implemented by photon exchange, 
which conserves the energy of the network. 
Unlike the first type, 
the second type of Ising machine can in principle be operated without dissipation 
and is based on quantum adiabatic evolution. 
Such a machine can be implemented with superconducting circuits, 
as suggested in \cite{Goto2016a}, 
and explicit circuit designs for all-to-all connectivity have been proposed 
in \cite{Nigg2017a,Puri2017a}.

In the present work, 
we numerically investigate the effects of dissipation on the second type of Ising machine with KPOs. 
Hereafter, we call this machine a \textit{quantum bifurcation machine}, or \textit{QbM} for short, 
because the operation principle is based on a quantum-mechanical bifurcation of the KPO network 
and is called bifurcation-based adiabatic quantum computation
\cite{Goto2016a}.
(We do not use ``QBM" because it is often used for quantum Boltzmann machine.) 
Our simulation results indicate 
that the probability distribution of the spin configurations in the dissipative QbM is Boltzmann-like 
with respect to the Ising energy. 
Similar phenomena have been observed in single driven dissipative quantum-mechanical nonlinear oscillators, 
where dissipation induces excitations in quasienergy levels. 
(The quasienergy levels are defined as eigenstates of the system Hamiltonian in a rotating frame.) 
This is called \textit{quantum heating} to distinguish it from thermal heating 
\cite{Dykman,Dykman2011a,Ong2013a}. 
We generalize the quantum heating to the case of multiple coupled oscillators 
and explain that the above Boltzmann distribution of the spin configurations is 
related to the generalized quantum heating. 
Although quantum heating causes errors in solving optimization problems, 
the Boltzmann distribution means that the QbM is robust against dissipation 
in the sense that good approximate solutions are obtained with high probability 
even in the presence of decoherence due to dissipation. 
(The robustness has recently been discussed from different points of view in \cite{Nigg2017a,Puri2017a}.) 
The present result also suggests that 
such driven dissipative nonlinear oscillator networks can be applied to \textit{Boltzmann sampling} 
from the Ising model. 
Recently, similar physical implementations of a Boltzmann sampler with Ising machines 
have also attracted much attention 
\cite{Dumoulin2014a,Adachi2015a,Benedetti2016a,Amin2015a,Amin2016a,Korenkevych2016a,Crawford2016a,Sakaguchi2016a} 
because it is useful for various purposes, 
such as Boltzmann machine learning in the field of artificial intelligence \cite{MacKay}.

This paper is organized as follows.
In Sec. \ref{sec-QbM},
the QbM is extended to the Ising problem with local fields,
and simulation results for the extended QbM in the absence of dissipation are shown.
In Sec. \ref{sec-simulation},
simulation results for dissipative QbMs are shown.
In Sec. \ref{sec-heating},
the results in Sec. \ref{sec-simulation} are discussed
from the viewpoint of quantum heating.
In Sec. \ref{sec-conclusion},
the present results are summarized and 
briefly disscussed with prospects for future work.

\section{Quantum bifurcation machine for the Ising problem with local fields}
\label{sec-QbM}

The QbM proposed in \cite{Goto2016a} only applies to the Ising problem \textit{without} local fields. 
In this paper, we extend the QbM to the Ising problem \textit{with} local fields, 
which is to find the spin configuration that minimizes the following dimensionless Ising energy:
\begin{align}
E_{\mbox{\scriptsize{Ising}}}(\vec{s})= 
-\frac{1}{2} \sum_{i=1}^N \sum_{j=1}^N J_{i,j} s_i s_j
+\sum_{i=1}^N h_i s_i,
\label{eq-Ising-energy}
\end{align}
where $s_i$ is the $i$-th Ising spin, 
which takes values of $+1$ (up) or $-1$ (down), 
$N$ is the total number of Ising spins, 
$\vec{s}=(s_1~s_2~\cdots~s_N)$ is the vector representation of a spin configuration, 
and $\{ J_{i,j} \}$ and $\{ h_i \}$ are the dimensionless parameters 
corresponding to the coupling coefficients and local fields, respectively. 
Note that $\{ J_{i,j} \}$ satisfies $J_{i,j}=J_{j,i}$ and $J_{i,i}=0$. 
This extension is significant because many applications such as the traveling salesman problem 
and Boltzmann machine learning require local fields \cite{Lucas2014a,Dumoulin2014a,Sakaguchi2016a,MacKay}.

For a given instance of the Ising problem, 
the extended QbM is defined by the following Hamiltonian in a frame rotating at half the pump frequency, 
$\omega_p/2$, of the parametric drive and in the rotating-wave approximation \cite{Goto2016a}:
\begin{align}
H
&=\hbar \sum_{i=1}^N \left[ \frac{K}{2} a_i^{\dagger 2} a_i^2 + \Delta a_i^{\dagger} a_i - \frac{p(t)}{2} \left( a_i^{\dagger 2} + a_i^2 \right) \right]
\nonumber \\
&-\hbar \xi_0 \sum_{i=1}^N \sum_{j=1}^N J_{i,j} a_i^{\dagger} a_j
+\hbar \xi_0 \alpha (t) \sum_{i=1}^N h_i \left( a_i^{\dagger} + a_i \right),
\label{eq-Hamiltonian}
\end{align}
where $a_i^{\dagger}$ and $a_i$ are the creation and annihilation operators for the $i$-th KPO, 
$K$ is the Kerr coefficient, 
$\Delta$ is the detuning frequency defined by $\Delta=\omega_{\mbox{\scriptsize{KPO}}}-\omega_p/2$ 
($\omega_{\mbox{\scriptsize{KPO}}}$ is the resonance frequency of the KPOs), 
$p(t)$ is the time-dependent pump amplitude, 
$\xi_0$ is a constant parameter with the dimension of frequency.
Here, we assume for simplicity that  $K$, $\Delta$, and $\xi_0$ are positive.
If $K$ is negative, as in the case of superconducting Josephson parametric oscillators \cite{Lin2014a}, 
we set $p(t)$, $\Delta$, and $\xi_0$ to negative values by flipping the signs. 
Then, we obtain the same result. 
The physical meaning of the third term, 
which is added for the extension, 
is the external drive of KPOs at $\omega_p/2$,
where $\xi_0 \alpha (t) h_i$ is the time-dependent amplitude of the external drive for the $i$-th KPO and 
$\alpha (t)$ is a dimensionless parameter defined such that 
$\alpha \approx 0$ when $p \ll \Delta$ and 
$\alpha \approx \sqrt{(p-\Delta)/K}$ when $p \gg \Delta$. 
Here, $\sqrt{(p-\Delta)/K}$ is the approximate magnitude of the amplitudes 
of the two stable oscillating states of each KPO \cite{Goto2016a}.

To find the ground state of the Ising model via quantum adiabatic evolution, we initialize all the KPOs in the ``vacuum" state 
and gradually increase the pump amplitude $p(t)$ from zero to a sufficiently large value 
compared to $\Delta$ and $\xi_0$. 
To satisfy the initial condition that the vacuum state is the ground state of the initial Hamiltonian, 
$\Delta$ is set such that the matrix $M$ defined by $M_{i,i}=\Delta$ and $M_{i,j}=-\xi_0 J_{i,j}$ ($i\neq j$) 
is positive semidefinite \cite{Goto2016a}. 
When $p(t)$ is increased, 
each KPO ends up approximately in either of two coherent states $|\pm \alpha (t) \rangle$. 
(A coherent state $|\alpha \rangle$ is defined as the eigenstate of an annihilation operator: 
$a_i |\alpha \rangle = \alpha |\alpha \rangle$ \cite{Leonhardt}.)
The expectation value of the Hamiltonian for the product of the coherent states 
$|\vec{s} \rangle := |s_1 \alpha \rangle |s_2 \alpha \rangle \cdots |s_N \alpha \rangle$ 
($s_i=\pm 1$ is the sign of the oscillation amplitude of the $i$-th KPO) is given by
\begin{align}
\langle \vec{s}|H| \vec{s} \rangle &=
\hbar 
\sum_{i=1}^N \left(
\frac{K}{2} \alpha^4 + \Delta \alpha^2 - p \alpha^2 \right)
\nonumber \\
&
+2\hbar \xi_0 \alpha^2
\left(
-\frac{1}{2} \sum_{i=1}^N \sum_{j=1}^N J_{i,j} s_i s_j
+\sum_{i=1}^N h_i s_i \right).
\label{eq-Hav}
\end{align}
Note that the first term is independent of $\{ s_i \}$ 
and the second term is proportional to the Ising energy $E_{\mbox{\scriptsize{Ising}}}(\vec{s})$ 
in Eq. (\ref{eq-Ising-energy}).

Assuming that the terms proportional to $\xi_0$ are small compared to the other terms in the Hamiltonian, 
the ground state minimizes $\langle \vec{s}|H| \vec{s} \rangle$ with respect to $\{ s_i \}$ 
by the perturbation theory to the lowest order \cite{Goto2016a}. 
Since the KPO network is kept in the instantaneous ground state of the time-dependent Hamiltonian during the adiabatic evolution, 
we obtain the state that minimizes $\langle \vec{s}|H| \vec{s} \rangle$. 
This state corresponds to the ground state of the given Ising model 
by identifying the sign of the quadrature amplitude defined by $x_i=(a_i + a_i^{\dagger})/2$ with the Ising spin $s_i$
\cite{Goto2016a}.

\begin{widetext}

\begin{figure}[htbp]
\begin{center}
	\includegraphics[width=14cm]{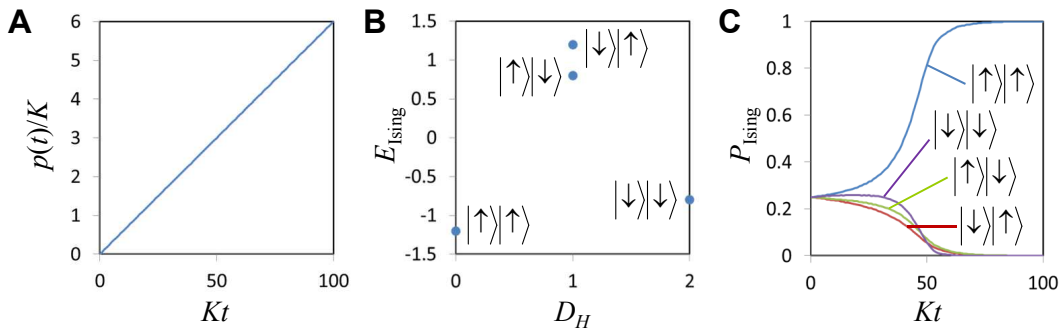}
	\caption{
	Nondissipative QbM. 
	(A) Time-dependent pump amplitude.
	(B) Energy landscape of an instance of the two-spin Ising model. 
	The parameters are $J_{1,2}=J_{2,1}=1$, $h_1=-0.2$, and $h_2=0$. 
	The horizontal axis represents the Hamming distance $D_H$ defined as 
	the number of spin flips with respect to the ground state $|\uparrow \rangle |\uparrow \rangle$. 
	(C) Time evolutions of the spin configuration probabilities $P_{\mbox{\scriptsize{Ising}}}(\vec{s})$ 
	given by Eq. (\ref{eq-PIsing}).}
	\label{fig1}
\end{center}
\end{figure}

\end{widetext}

To verify the validity of the above discussion, 
we numerically investigate an instance of two KPOs ($N=2$), where
the Schr\"{o}dinger equation with the Hamiltonian in Eq. (\ref{eq-Hamiltonian}) is solved numerically
and the time-dependent pump amplitude $p(t)$ is increased linearly, as shown in Fig. \ref{fig1}A.
The parameters of the instance are $J_{1,2}=J_{2,1}=1$, $h_1=-0.2$, and $h_2=0$, which 
are set such that two local minima exist in the energy landscape, 
as shown in Fig. \ref{fig1}B.

In the present numerical study, the Hilbert space is truncated at a ``photon" number of 14 for each KPO
and 
$\alpha (t)$ is set to the following form that satisfies the above conditions:
\begin{align}
\alpha (t) = \sqrt{\frac{p(t)-\Delta \tanh [p(t)/\Delta]}{K}}.
\label{eq-alpha}
\end{align}
The other parameters are set to $\Delta =2K$ and $\xi_0=0.5K$.

Figure \ref{fig1}C shows the time evolutions of the spin configuration probabilities 
$P_{\mbox{\scriptsize{Ising}}}(\vec{s})$ given by
\begin{align}
P_{\mbox{\scriptsize{Ising}}}(\vec{s})
=
\mbox{Tr}
\left[
\rho \prod_{i=1}^N M^{(i)}(s_i)
\right],
\label{eq-PIsing}
\end{align}
where $\rho$ is the density operator describing the state of the system and
\begin{align}
M^{(i)}(1) = \int_0^{\infty} dx_i |x_i \rangle \langle x_i|,
M^{(i)}(-1) = \int^0_{-\infty} dx_i |x_i \rangle \langle x_i|
\label{eq-POVM}
\end{align}
compose the positive-operator-valued measure 
for measuring the sign of the quadrature amplitude, $x_i$, of the $i$-th KPO
($|x_i \rangle$ is the eigenstate of $x_i$). 
The probabilities $P_{\mbox{\scriptsize{Ising}}}(\vec{s})$ are calculated by the method presented in \cite{Goto2016a}.
As shown in Fig. \ref{fig1}C, 
the state of the two KPOs finally converges to $|\uparrow \rangle |\uparrow \rangle$, 
which is the ground state of the given Ising model, as expected.

\section{Dissipative quantum bifurcation machine}
\label{sec-simulation}

In the presence of dissipation, 
the time evolution of a QbM is modeled by the following quantum master equation 
\cite{Dykman,Dykman2011a,Breuer}:
\begin{align}
\dot{\rho}
=
&-\frac{i}{\hbar} [H,\rho]
+\kappa (\bar{n}+1) \sum_{i=1}^N \left(
2a_i \rho a_i^{\dagger}
-a_i^{\dagger} a_i \rho
-\rho a_i^{\dagger} a_i
\right)
\nonumber \\
&+\kappa \bar{n} \sum_{i=1}^N \left(
2a_i^{\dagger} \rho a_i
-a_i a^{\dagger}_i \rho
-\rho a_i a^{\dagger}_i
\right),
\label{eq-master}
\end{align}
where the dot denotes the time derivative, 
$\kappa$ is the decay rat of the KPOs characterizing the dissipation,
and $\bar{n}=\{ \exp [ \hbar (\omega_p/2) /(k_B T)] -1\}^{-1}$ is the Planck number  at frequency $\omega_p/2$ 
and temperature $T$ ($k_B$ is the Boltzmann constant). 
While the first term in the right-hand side of Eq. (\ref{eq-master}) describes 
the unitary time evolution of the system, the other terms are for the non-unitary evolution.
In the following, $\bar{n}$ is set to zero assuming a sufficiently low temperature.

We numerically solve the master equation for the same instance as above. 
In the simulations of dissipative QbMs, we use the following form of $p(t)$:
\begin{align}
p(t)=p_f \tanh (3t/\tau),
\label{eq-pt}
\end{align}
where $p_f$ is the final value of $p(t)$ and 
$\tau$ is the time at which $p(t)$ closely approaches $p_f$. 
The form of $p(t)$ is chosen such that $p(t)$ increases linearly with respect to $t$ 
at the initial time and converges to its final value $p_f$. 
The time $\tau$ is set to $100/K$ in the present work.

\begin{widetext}

\begin{figure}[htbp]
\begin{center}
	\includegraphics[width=14cm]{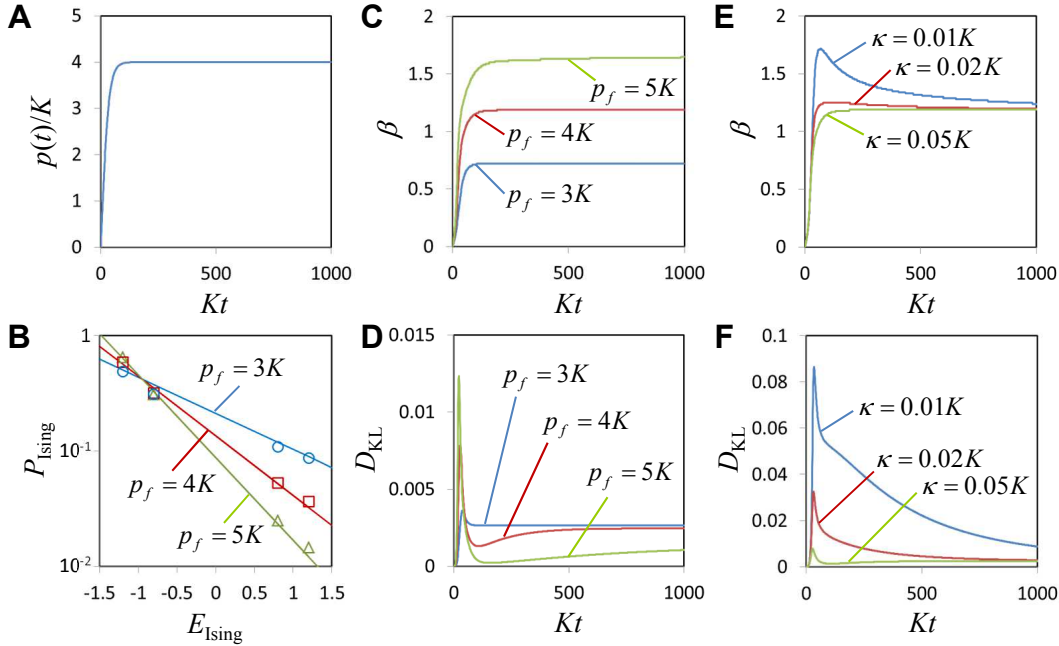}
	\caption{
	Dissipative QbM. 
	(A) Time-dependent pump amplitude $p(t)$ [Eq. (\ref{eq-pt}) with $p_f=4K$]. 
	(B) Probability distributions of the spin configurations. 
	Symbols show $P_{\mbox{\scriptsize{Ising}}}(\vec{s})$ at the final time ($Kt=1000$) 
	obtained with the numerical solution of 
	the quantum master equation (\ref{eq-master}). 
	Circles, squares, and triangles correspond to $p_f=3K$, $4K$, and $5K$, respectively.
	The decay rate is set to $\kappa =0.05K$.
	The lines show the Boltzmann distribution fitting to the simulation results. 
	(C) Inverse effective temperature $\beta$ for the three values of $p_f$ 
	determined by fitting to the instantaneous probability distribution. 
	(D) Kullbak-Leibler (KL) divergence $D_{\mbox{\scriptsize{KL}}}$ minimized for the fitting in (C). 
	(E and F) Time evolutions of $\beta$ and $D_{\mbox{\scriptsize{KL}}}$ 
	for various values of $\kappa$ ($p_f=4K$).}
	\label{fig2}
\end{center}
\end{figure}

\end{widetext}

The simulation results are summarized in Fig. \ref{fig2}. 
Figure \ref{fig2}A shows the time-dependent pump amplitude $p(t)$ with $p_f=4K$. 
The symbols in Fig. \ref{fig2}B shows the distribution of the spin configuration probabilities 
$P_{\mbox{\scriptsize{Ising}}}(\vec{s})$
with respect to the Ising energy $E_{\mbox{\scriptsize{Ising}}}(\vec{s})$ at the final time ($Kt=1000$), 
where the decay rate is set to $\kappa=0.05K$. 
The line in Fig. Fig. \ref{fig2}B is obtained by fitting the Boltzmann distribution to the simulation results, 
where the Kullback-Leibler (KL) divergence $D_{\mbox{\scriptsize{KL}}}$ 
between the two distributions is minimized (see Appendix \ref{Appendix} for details). 
The Boltzmann distribution is defined by 
$\displaystyle P_B(\vec{s},\beta)=\exp [-\beta E_{\mbox{\scriptsize{Ising}}}(\vec{s})]/Z(\beta)$, 
where $\beta$ is the inverse effective temperature and 
$\displaystyle Z(\beta)=\sum_{\vec{s}} \exp [-\beta E_{\mbox{\scriptsize{Ising}}}(\vec{s})]$ is the partition function. 
In the fitting, $\beta$ is a single fitting parameter. 
The good fits shown in Fig. \ref{fig2}B indicate 
that the probability distributions of the spin configurations in the dissipative QbM are Boltzmann-like.

Figures \ref{fig2}C and \ref{fig2}D show the time evolutions of the inverse effective temperature $\beta$ 
and the minimized KL divergence $D_{\mbox{\scriptsize{KL}}}$, respectively.
Figure 2C shows that as $p_f$ increases, $\beta$ increases, that is, the effective temperature decreases. 
Thus, the effective temperature can be controlled by the pump amplitude of the parametric drive.

Figures \ref{fig2}E and \ref{fig2}F show similar results for different values of $\kappa$ ($p_f =4K$).
Figure 2E shows that $\beta$ converges to a single value independent of $\kappa$, and
Fig. \ref{fig2}F supports that the probability distributions approach the Boltzmann distribution.

\begin{figure}[htb]
\begin{center}
	\includegraphics[width=8.5cm]{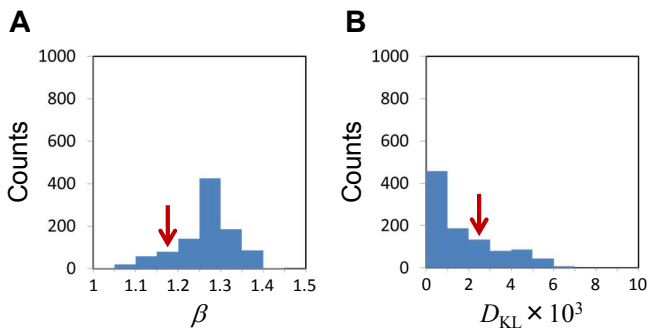}
	\caption{Simulation results of 1000 instances of the two-spin Ising problem. 
	$\{ J_{i,j} \}$ and $\{ h_i \}$ are chosen randomly from the interval $(-1, 1)$. 
	The other parameters are set to $p_f=4K$ and $\kappa =0.05K$. 
	(A) Histogram of the inverse effective temperature $\beta$ determined 
	by fitting to the final probability distribution. 
	(B) Histogram of the corresponding Kullbak-Leibler (KL) divergence $D_{\mbox{\scriptsize{KL}}}$ 
	minimized for the fitting in (A). 
	The arrows indicate the results of the instance in Fig. \ref{fig2}.}
	\label{fig3}
\end{center}
\end{figure}

To check that the probability distribution of the spin configurations is 
also Boltzmann-like for other instances, 
we perform similar numerical simulations for 1000 instances of the two-spin Ising problem,
where their parameters, $\{ J_{i,j} \}$ and $\{ h_i \}$, 
are chosen randomly from the interval $(-1, 1)$. 
The other parameters are set to $p_f=4K$ and $\kappa =0.05K$. 
The results for $\beta$ and $D_{\mbox{\scriptsize{KL}}}$ are shown in Fig. \ref{fig3}, 
where the arrows indicate the results of the instance shown in Fig. \ref{fig2}. 
The averages and standard deviations are $\beta = 1.27 \pm 0.07$ and 
$D_{\mbox{\scriptsize{KL}}}=(1.7 \pm 1.7) \times 10^{-3}$. 
The largest value of $D_{\mbox{\scriptsize{KL}}}$ is $6.5 \times 10^{-3}$. 
On the other hand, we obtain $D_{\mbox{\scriptsize{KL}}}=(2.0 \pm 2.3) \times 10^{-1}$ 
when the spin configuration probabilities for each instance are set randomly 
by choosing four random numbers from the interval $(0, 1)$ and normalizing them. 
This comparison shows that the probability distributions of the spin configurations 
in the thousand cases are Boltzmann-like compared to general distributions. 
Note also that the instance dependence of $\beta$ is small 
in the sense that the standard deviation is much smaller than the average.

We also simulate an instance of the four-spin Ising problem 
to check the case with more than two spins ($N>2$). 
Since it is computationally hard to solve the quantum master equation 
in the four-spin case, we use the quantum-jump approach 
\cite{Breuer,Plenio1998a}, 
which is a Monte-Carlo simulation using a state vector, 
instead of a density matrix, and can provide equivalent results to the quantum master equation. 
The probability distribution obtained is also Boltzmann-like. 
(See Appendix \ref{sec-four-spin} for details.)

\section{Quantum heating in dissipative QBM}
\label{sec-heating}

we first generalize the quantum heating to the case of multiple coupled nonlinear oscillators, 
and then explain that the above Boltzmann distribution of the spin configurations is 
related to the generalized quantum heating.

Quantum heating is the heating process induced by quantum jumps due to dissipation in quasienergy levels of a driven system. 
This is well described by the balance equation derived from the quantum master equation \cite{Dykman}. 
We apply the balance-equation approach to the case of multiple coupled oscillators.

Using the quasienergy states [eigenstates of the Hamiltonian in Eq. (\ref{eq-Hamiltonian})], 
$\{ |E_n \rangle \}$, as an orthonormal basis, 
the master equation (\ref{eq-master}) becomes a system of ordinary differential equations of the density matrix 
$\rho_{m,n} = \langle E_m|\rho |E_n \rangle$. 
Note that in the equations for the diagonal elements $\{ \dot{\rho}_{n,n} \}$, 
the terms for the unitary time evolution are cancelled out.
By disregarding the off-diagonal elements, we obtain the following balance equation 
with respect to the diagonal elements \cite{Dykman}:
\begin{align}
\dot{\rho}_{n,n}=
2\kappa \sum_{i=1}^N \sum_{m=0}^{\infty} \left( |a_{n,m}^{(i)}|^2 \rho_{m,m}-|a_{m,n}^{(i)}|^2 \rho_{n,n} \right),
\label{eq-balance}
\end{align}
where $a_{m,n}^{(i)}=\langle E_m|a_i |E_n \rangle$.
Note that the diagonal element $\rho_{n,n}$ represents the probability that the system is in the quasienergy state $|E_n \rangle$.

A physical interpretation of the balance equation (\ref{eq-balance}) is as follows. 
Dissipation induces quantum jumps corresponding to one-photon loss \cite{Breuer,Plenio1998a}. 
A quantum jump by an annihilation operator $a_i$ changes $|E_n \rangle$ into $a_i|E_n \rangle$, 
and consequently causes the transition from $|E_n \rangle$ to $|E_m \rangle$ with probability proportional to $|\alpha_{m,n}^{(i)}|^2$. 
Quantum heating is the heating process that originates from the transitions due to quantum jumps.
Note that coherent states, which are often regarded as classical states, 
are eigenstates of annihilation operators, and therefore 
the transitions due to quantum jumps do not occur for coherent states.

\begin{widetext}

\begin{figure}[htb]
\begin{center}
	\includegraphics[width=14cm]{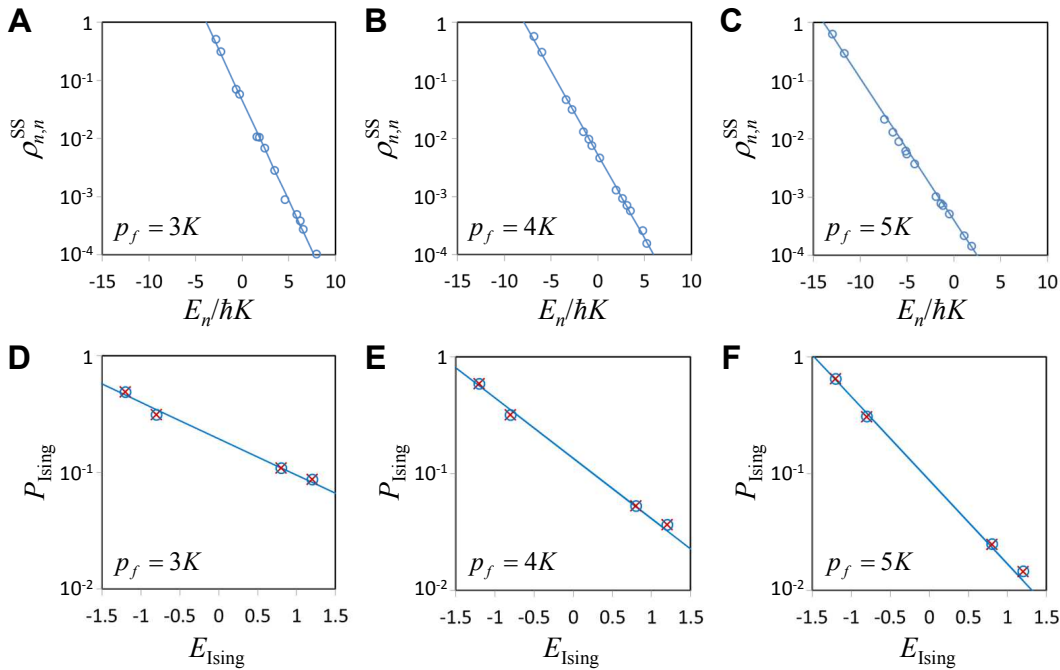}
	\caption{Quantum heating of dissipative QbM. 
	(A--C) Probability distribution, $\rho^{\mbox{\scriptsize{SS}}}_{n,n}$, 
	of quasienergies, $E_n$, for the steady-state solution of the balance equation (\ref{eq-balance}). 
	Circles show numerical results and the lines show exponential fits. 
	(D--F) Comparison between $P^{\mbox{\scriptsize{BE}}}_{\mbox{\scriptsize{Ising}}}(\vec{s})$ 
	given by Eq. \ref{eq-PBE} (crosses) and $P_{\mbox{\scriptsize{Ising}}}(\vec{s})$ 
	given by Eq. \ref{eq-PIsing} (circles). The lines show the Boltzmann distribution fitting to 
	$P_{\mbox{\scriptsize{Ising}}}(\vec{s})$.}
	\label{fig4}
\end{center}
\end{figure}

\end{widetext}

The steady-state solution, $\{ \rho^{\mbox{\scriptsize{SS}}}_{n,n} \}$, 
of the balance equation is obtained by substituting $\dot{\rho}_{n,n}=0$ into Eq. (\ref{eq-balance}) 
under the constraint $\displaystyle \sum_{n=1}^{\infty} \rho_{n,n}=1$. 
We numerically evaluate the steady-state solution for the instance in Fig. \ref{fig2}. 
The results are given in Figs. \ref{fig4}A--\ref{fig4}C. 
The probability distributions are clearly Boltzmann-like. 
This result suggests that the quasienergies of coupled nonlinear oscillators 
obey the Boltzmann distribution due to quantum heating.

The spin configuration probabilities, $P^{\mbox{\scriptsize{BE}}}_{\mbox{\scriptsize{Ising}}}(\vec{s})$, 
for the steady state of the balance equation are given by
\begin{align}
P^{\mbox{\scriptsize{BE}}}_{\mbox{\scriptsize{Ising}}}(\vec{s})
=
\sum_n \rho^{\mbox{\scriptsize{SS}}}_{n,n} P^{(n)}_{\mbox{\scriptsize{Ising}}}(\vec{s}),
\label{eq-PBE}
\end{align}
where $P^{(n)}_{\mbox{\scriptsize{Ising}}}(\vec{s})$ represent the spin configuration probabilities 
for the quasienergy state $|E_n \rangle$, that is, 
\begin{align}
P^{(n)}_{\mbox{\scriptsize{Ising}}}(\vec{s})
=
\langle E_n| \prod_{i=1}^N M^{(i)}(s_i) |E_n \rangle.
\label{eq-Pn}
\end{align}
The comparison between $P^{\mbox{\scriptsize{BE}}}_{\mbox{\scriptsize{Ising}}}(\vec{s})$
and $P_{\mbox{\scriptsize{Ising}}}(\vec{s})$ in Fig. \ref{fig2}B is 
shown in Figs. \ref{fig4}D--\ref{fig4}F. 
They are in excellent agreement with each other. 
Hence, the Boltzmann distribution of the spin configurations is well explained 
by the generalized quantum heating. 
This result can be understood under some approximations as follows. 
From the generalized quantum heating, 
the density operator is approximately given by 
$\rho^{\mbox{\scriptsize{SS}}} =\exp (-\beta' H)/Z'(\beta')$, 
where $\beta'$ is the inverse effective temperature and 
$Z'(\beta') = \mbox{Tr}[\exp (-\beta' H)]$ is the corresponding partition function. 
(The primes are used to distinguish them from the above ones for the spin configurations.) 
When the dissipation is sufficiently small, 
the state is approximately one of the stable oscillating states 
$|\vec{s} \rangle := |s_1 \alpha \rangle |s_2 \alpha \rangle \cdots |s_N \alpha \rangle$ 
($s_i=\pm 1$). 
By the classical approximation that the annihilation operator $a_i$ is 
replaced by the amplitude $s_i\alpha$, 
we obtain [also see Eq. (\ref{eq-Hav})]
\begin{align}
P^{\mbox{\scriptsize{BE}}}_{\mbox{\scriptsize{Ising}}}(\vec{s})
\approx
\langle \vec{s}|\rho^{\mbox{\scriptsize{SS}}}|\vec{s} \rangle
\propto
\exp \left[
-2\hbar \xi_0 \alpha^2 \beta' E_{\mbox{\scriptsize{Ising}}}(\vec{s})
\right].
\label{eq-QH-approx}
\end{align}
Thus, the quantum heating leads to the Boltzmann distribution of the spin configurations. 
Moreover, this derivation indicates that $\beta=2\hbar \xi_0 \alpha^2 \beta'$. 
In the case of Fig. \ref{fig4}, 
the ratio $2\hbar \xi_0 \alpha^2 \beta'/\beta$ is close to unity, 
as expected (the values are 1.11, 1.12, and 1.02 for $p_f=3K$, $4K$, and $5K$, respectively). 
This supports the above derivation.

It is also notable that the steady-state solution of the balance equation (\ref{eq-balance}) 
is independent of $\kappa$. 
This can explain why $\beta$ in Fig. \ref{fig2}E converges to a single value independent of $\kappa$. 
The decay-rate-independent $\beta$ is a feature of quantum heating 
\cite{Dykman2011a}.

\section{Conclusions}
\label{sec-conclusion}

We have found by numerical simulation 
that the probability distributions of the spin configurations in dissipative QbMs are Boltzmann-like. 
We have also explained that the Boltzmann distribution is 
related to the quantum heating generalized to multiple coupled nonlinear oscillators. 
The present work is based on numerical analysis. 
Further general and analytic treatment is desirable in future work.

It is expected to be feasible for current technologies 
to experimentally observe the quantum heating of a driven dissipative nonlinear oscillator network. 
The most promising physical system for this is superconducting circuits 
because they have already been used for the experiments on quantum heating of a single nonlinear oscillator \cite{Ong2013a}, 
parametric oscillations \cite{Lin2014a}, and large Kerr effects \cite{Kirchmair2013a,Rehak2014a}. 
It is also notable that the Boltzmann distribution of the spin configurations 
can be observed by the measurement of quadrature amplitudes with heterodyne detection \cite{Lin2014a}, 
which is easier than the measurement of quasienergy states.

The present result also broadens the potential applications of QbM, 
such as Boltzmann sampling, which is used, e.g., for Boltzmann machine learning. 
Although it is an important and intriguing question 
whether or not the Boltzmann sampling using the dissipative QbM has some speedup over classical algorithms, 
this is beyond the scope of this paper. 
Nevertheless, our proposal is expected to open a new possibility 
for harnessing the behaviors of complex open quantum systems for practical applications. 
Thus, the present work is expected to trigger interdisciplinary research in the fields of quantum information science, 
nonequilibrium quantum systems, nonlinear dynamics, and artificial intelligence.

\section*{Acknowledgments}

This work was supported by JST ERATO (Grant No. JPMJER1601).

\begin{appendix}

\section{Fitting of the Boltzmann distribution to the simulation results}
\label{Appendix}

Here, 
the Boltzmann distribution is fitted to the simulation results 
by minimizing the Kullback-Leibler (KL) divergence $D_{\mbox{\scriptsize{KL}}}$, 
where the KL divergence between two probability distributions $\{ P_n \}$ and $\{ Q_n \}$ is defined as follows 
\cite{MacKay}:
\begin{align}
D_{\mbox{\scriptsize{KL}}} (P||Q)=
\sum_n P_n \ln \frac{P_n}{Q_n}.
\label{eq-KL}
\end{align}
Note that the KL divergence is asymmetric with respect to the two distributions. 
In this paper, we choose the Boltzmann distribution as $\{ P_n \}$ and the simulation results as $\{ Q_n \}$.

\section{Four-spin Ising problem}
\label{sec-four-spin}

Here, we provide simulation results for an instance of the four-spin Ising problem 
in order to check whether the probability distribution of the spin configurations 
in the dissipative QbM is also Boltzmann-like in the case with more than two spins ($N>2$). 
As mentioned in the main text, 
it is computationally hard to solve the quantum master equation in the four-spin case. 
We therefore use the quantum-jump approach \cite{Breuer,Plenio1998a}.

\begin{widetext}

\begin{figure}[htbp]
\begin{center}
	\includegraphics[width=10cm]{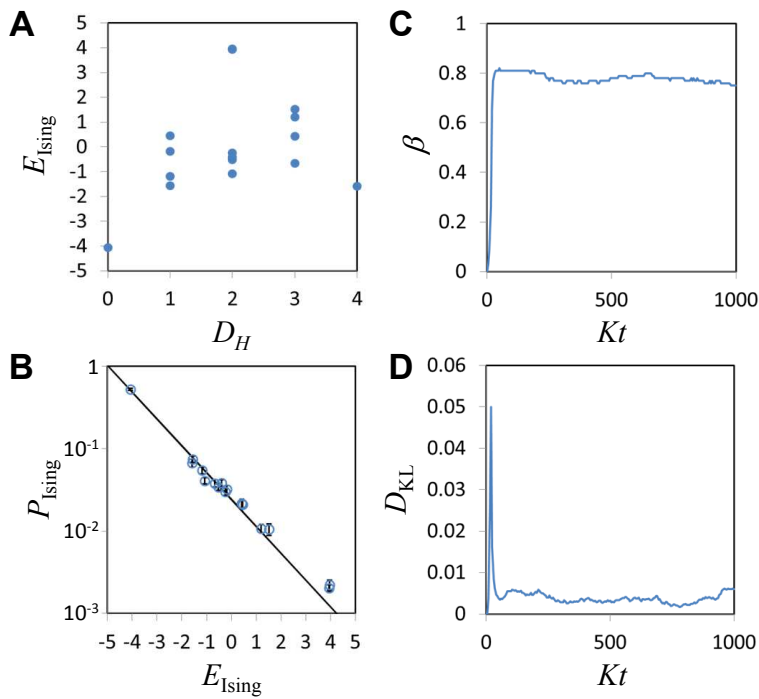}
	\caption{
	Four-spin Ising model. 
	(A) Energy landscape of the instance. 
	The parameters are given in the text. 
	(B) Distribution of the spin configuration probabilities $P_{\mbox{\scriptsize{Ising}}}(\vec{s})$ 
	at the final time ($Kt=1000$). 
	Error bars represent standard errors. 
	The line shows the Boltzmann distribution fitting to the simulation results. 
	(C) Inverse effective temperature $\beta$ determined by fitting to the instantaneous probability distribution. 
	(D) Kullbak-Leibler (KL) divergence $D_{\mbox{\scriptsize{KL}}}$ minimized for the fitting in (C).}
	\label{figS1}
\end{center}
\end{figure}

\end{widetext}

The parameters are given by 
$J_{1,2}=J_{2,1}=0.93406$, 
$J_{1,3}=J_{3,1}=0.801243$, 
$J_{1,4}=J_{4,1}=0.094465$, 
$J_{2,3}=J_{3,2}=-0.654609$, 
$J_{2,4}=J_{4,2}=0.945369$, 
$J_{3,4}=J_{4,3}=0.711242$, 
$h_{1}=0.429632$,
$h_{2}=0.218071$,
$h_{3}=0.395458$,
$h_{4}=0.195112$,
which were chosen randomly from the interval $(-1, 1)$. 
The energy landscape of this instance is shown in Fig. \ref{figS1}A. 
The pump amplitude $p(t)$ follows Eq. (\ref{eq-pt}) with $p_f=6K$. 
The other parameters are set to $\Delta =2K$, $\xi_0=0.2K$, and $\kappa =0.02K$.

The simulation results are summarized in Fig. \ref{figS1}. 
The spin configuration probabilities $P_{\mbox{\scriptsize{Ising}}}(\vec{s})$ 
are obtained by taking the average over 300 trajectories of the Monte-Carlo simulation. 
As shown in Fig. \ref{figS1}B, the probability distribution is Boltzmann-like.

\end{appendix}

\end{document}